\documentclass[nofootinbib,showpacs,preprintnumbers,amsmath,amssymb,floatfix]
{revtex4}
\usepackage{epsf}

\begin{document}


\title{Deep Inelastic Spin Structure Functions at small x }

\vspace*{0.3 cm}

\author{B.I.~Ermolaev}
\affiliation{Ioffe Physico-Technical Institute, 194021
  St.Petersburg, Russia}
\author{M.~Greco}
\affiliation{Department of Physics and INFN, University Rome III,
Rome, Italy}
\author{S.I.~Troyan}
\affiliation{St.Petersburg Institute of Nuclear Physics, 188300
Gatchina, Russia}

\begin{abstract}
Explicit expressions for the non-singlet and singlet spin-dependent structure function $g_1$
in the  small-$x$ region are obtained. They include the total resummation  of the double- and single- logarithms 
of $x$ and account for the running QCD coupling $\alpha_s$ effects. Both the non-singlet and singlet
structure functions are Regge behaved asymptotically, with the intercepts predicted in agreement with
 experiments. A detailed comparison with the DGLAP evolution equations for different values of $x$ and 
 $Q^2$ is performed. Finally, the role played by singular terms in 
 DGLAP fits for the initial quark densities is discussed and  
 explicitly shown to mimic the resummation of leading logarithms at
 small-$x$.
\end{abstract}


\maketitle

\section{Introduction}

In the standard theoretical approach  for investigating the DIS
structure function $g_1(x, Q^2)$, namely DGLAP\cite{dglap},
 $g_1^{DGLAP}$ is a convolution of the coefficient
functions $C_{DGLAP}$ and the evolved parton distributions. The latter 
 are also expressed as a convolution of the splitting functions
$P_{DGLAP}$ and initial parton densities,  which are fitted
from experimental data at large $x$, $x \sim 1$ and  $Q^2 \sim
1$~GeV$^2$. However there is an obvious asymmetry in treating 
the $Q^2$- and $x$- logarithmic contributions in DGLAP.
 Indeed, the leading $Q^2$-
contributions, $\ln(Q^2)$, are accounted to all orders in
$\alpha_s$ whereas $C_{DGLAP}(x)$ and $P_{DGLAP}(x)$ are known in
first two orders of the perturbative QCD. 
On the other hand,
in the small-$x$ region the situation looks opposite: logarithms
of $x$, namely double logarithmis(DL), i.e. the terms $~(\alpha_s
\ln^2(1/x))^k$, and single logarithms (SL), the terms $~(\alpha_s
\ln(1/x))^k$,with $k
 = 1,2,..$, are becoming quite sizable and should be accounted to all ordersi
in $\alpha_s$.  The total resummation of DL terms was first
done\cite{ber} in the fixed  $\alpha_s$ approximation, and  led to a new expression
$g_1^{DL}$, for $g_1$, that in the small-$x$
asymptotics  was of the Regge (power-like) form and
much greater than the well-known small-$x$ asymptotics of
$g_1^{DGLAP}$. 

Strictly speaking, the results of Refs.~\cite{ber} 
 could not be compared in a straightforward way with DGLAP because
instead of  the running $\alpha_s$,  with the
parametrization
\begin{equation}\label{dglapparam}
\alpha_s^{DGLAP} = \alpha_s(Q^2),
\end{equation}
Refs.~\cite{ber} had used $\alpha_s$ fixed at an  unknown
scale.  A closer investigation of  this matter\cite{egt1}
led us to conclude that the DGLAP- parametrization of
Eq.~(\ref{dglapparam}) can be a good approximation at $x$ not far
from $1$ only. Instead, a new parametrization was suggested,
where the argument of $\alpha_s$ in each of the ladder rungs of the 
involved
Feynman diagram is the
virtuality of the horizontal gluon (see Ref.~\cite{egt1} for
detail). 
Indeed this parametrization works  well both for small
and large $x$.  It converges to the DGLAP- parametrization at
large $x$ but differs from it at small $x$, and it
  allowed us to obtain in Refs.~\cite{egt2} the
expressions for $g_1$ accounting for all-order resuummations of DL
and SL terms, including the running $\alpha_s$ effects
\footnote{The parametrization of see Ref.~\cite{egt1} was used
later in Refs.~\cite{kotl} for studying the small-$x$ contribution
to the Bjorken sum rule.}.  This led us  to predict
the numerical values of the intercepts of the singlet and
non-singlet $g_1$. These results were
then confirmed\cite{kat} by several independent groups who 
have analyzed the HERMES data and extrapolated them at small $x$.

On the other hand, it is well known that, despite missing the total
resummation of $\ln x$, DGLAP works quite successfully  at $x \ll 1$.
This might suggest that the total
resummation of DL contributions performed in Refs.~\cite{egt2}
should not be relevant at available  values of $x$ and might be of some
importance at extremely small $x$ reachable in the future. In
Ref.~\cite{egt3} we made a detailed numerical analysis and
explained why DGLAP fits can be successful at small $x$. Indeed in order 
to describe the available experimental data,
singular expressions (see for example Refs.~\cite{a,v}) are introduced for the
initial parton densities. These singular factors (i.e. the factors which
$\to \infty$ when $x \to 0$ ) introduced in the fits mimic the total
resummaton of Refs.~\cite{egt2}. Then  using the results of
Ref~\cite{egt2} for incorporating the total resummation of $\ln x$
it allows to simplify the rather sophisticated
structure of the standard DGLAP fits down to a normalization
constant at small $x$.

\section{Difference between DGLAP and our approach}

 In DGLAP, $g_1$ is expressed through convolutions of the coefficient functions
and evolved parton distributions. As convolutions look simpler in
terms of integral transforms, it is convenient to represent $g_1$
in the form of the Mellin integral. For example, the non-singlet
component  of $g_1$ can be represented as follows:

\begin{equation}
\label{fdglapmellin} g^{NS}_{1~DGLAP}(x, Q^2) = (e^2_q/2)
\int_{-\imath \infty}^{\imath \infty} \frac{d \omega}{2\imath
\pi}(1/x)^{\omega} C_{DGLAP}(\omega) \delta q(\omega) \exp \Big[
\int_{\mu^2}^{Q^2} \frac{d k^2_{\perp}}{k^2_{\perp}}
\gamma_{DGLAP}(\omega, \alpha_s(k^2_{\perp}))\Big]
\end{equation}
with $C_{DGLAP}(\omega)$ being the non-singlet coefficient
functions, $\gamma_{DGLAP}(\omega,  \alpha_s)$ the non-singlet
anomalous dimensions and $\delta q(\omega)$ the initial
non-singlet quark densities in the  Mellin (momentum) space. The
expression for the singlet $g_1$ is similar, though more involved.
Both $\gamma_{DGLAP}$ and $C_{DGLAP}$  are known in first two
orders of the perturbative QCD. Technically, it is simpler to
calculate them at integer values of $\omega = n$. In this case,
the
 integrand of Eq.~(\ref{fdglapmellin})  is called the $n$-th momentum of
 $g^{NS}_1$. Once the moments for different $n$ are known, $g^{NS}$ at arbitrary values of $\omega$
 is obtained by interpolation. 
 The expressions of the initial quark densities are obtained from phenomenological consideration,
 by fitting the experimental data at $x \sim 1$.  Eq.~(\ref{fdglapmellin})  shows that $\gamma_{DGLAP}$
 govern the $Q^2$- evolution whereas  $C_{DGLAP}$ evolve
 $\delta q(\omega)$
 in the $x$-space from $x \sim 1$ into the small $x$ region.
 When in the $x$-space the initial parton distributions $\delta
q(x)$ are regular in $x$, i.e. do not $\to \infty$ when $x \to 0$,
the small-$x$ asymptotics of $g_{1~DGLAP}$ is given by the
well-known expression:

\begin{equation}\label{dglapas}
 g^{NS}_{1~DGLAP},~~g^{S}_{1~DGLAP} \sim \exp \Big[ \sqrt{\ln (1/x) \ln
 \Big( \ln(Q^2/\mu^2)/\ln (\mu^2/
 \Lambda^2_{QCD})\Big)}~\Big].
\end{equation}

On the contrary, when the total resummation of the double-logarithms (DL) and single-
logarithms of $x$ is done\cite{egt1}, the Mellin representation
for $g_1^{NS}$  is
\begin{equation}
\label{gnsint} g_1^{NS}(x, Q^2) = (e^2_q/2) \int_{-\imath
\infty}^{\imath \infty} \frac{d \omega}{2\pi\imath }(1/x)^{\omega}
C_{NS}(\omega) \delta q(\omega) \exp\big( H_{NS}(\omega)
\ln(Q^2/\mu^2)\big)~,
\end{equation}
with new coefficient functions  $C_{NS}$,
\begin{equation}
\label{cns} C_{NS}(\omega) =\frac{\omega}{\omega -
H_{NS}^{(\pm)}(\omega)}
\end{equation}
and anomalous dimensions $H_{NS}$,
\begin{equation}
\label{hns} H_{NS} = (1/2) \Big[\omega - \sqrt{\omega^2 -
B(\omega)} \Big]
\end{equation}
where
\begin{equation}
\label{b} B(\omega) = (4\pi C_F (1 +  \omega/2) A(\omega) +
D(\omega))/ (2 \pi^2)~.
\end{equation}
 $ D(\omega)$ and $A(\omega)$ in Eq.~(\ref{b}) are
expressed in terms of  $\rho = \ln(1/x)$, $\eta =
\ln(\mu^2/\Lambda^2_{QCD})$, $b = (33 - 2n_f)/12\pi$ and the color
factors
 $C_F = 4/3$, $N = 3$:

\begin{equation}
\label{d} D(\omega) = \frac{2C_F}{b^2 N} \int_0^{\infty} d \rho
e^{-\omega \rho} \ln \big( \frac{\rho + \eta}{\eta}\big) \Big[
\frac{\rho + \eta}{(\rho + \eta)^2 + \pi^2} \mp
\frac{1}{\eta}\Big] ~,
\end{equation}

\begin{equation}
\label{a} A(\omega) = \frac{1}{b} \Big[\frac{\eta}{\eta^2 + \pi^2}
- \int_0^{\infty} \frac{d \rho e^{-\omega \rho}}{(\rho + \eta)^2 +
\pi^2} \Big].
\end{equation}
$H_{S}$  and $C_{NS}$ account for DL and SL contributions to all
orders in $\alpha_s$. 

When $x \to 0$,
\begin{equation}
\label{gnsas}g_1^{NS} \sim \big( x^2/Q^2\big)^{\Delta_{NS}/2},~
g_1^{S} \sim \big( x^2/Q^2\big)^{\Delta_{S}/2}
\end{equation}
where the non-singlet and singlet intercepts are $\Delta_{NS} =
0.42,~\Delta_{S} = 0.86$. The $x$- behaviour of Eq.~(\ref{gnsas})
is much steeper than the one of Eq.~(\ref{dglapas}). 
Obviously, the total resummation of logarithms of $x$ leads to a faster
growth of $g_1$ when $x$ is decreasing, compared to what is predicted
by DGLAP, provided the input initial parton density $\delta q$ in
Eq.~(\ref{fdglapmellin}) is a regular function of $\omega$ at
$\omega \to 0$.

\section{Role of the initial parton densities}
There are various forms in the literature for  $\delta q(x)$, 
but all available fits include  both a regular and a singular
factor when $x \to 0$ (see e.g. Refs.~\cite{a,v} for detail). For
example, one of the fits from Ref.~\cite{a} is given by
\begin{equation}
\label{fita} \delta q(x) = N \eta x^{- \alpha} \Big[(1
-x)^{\beta}(1 + \gamma x^{\delta})\Big],
\end{equation}
with $N,~\eta$ being normalization factors, $\alpha = 0.576$, $\beta =
2.67$, $\gamma = 34.36$ and $\delta = 0.75$. In the $\omega$
-space Eq.~(\ref{fita}) is a sum of pole contributions:
\begin{equation}
\label{fitaomega} \delta q(\omega) = N \eta \Big[ (\omega -
\alpha)^{-1} + \sum m_k (\omega + \lambda_k)^{-1}\Big],
\end{equation}
with $\lambda_{k} > 0$, and the first term in
Eq.~(\ref{fitaomega}) corresponds to the singular factor
$x^{-\alpha}$ of Eq.~(\ref{fita}). When  Eq.~(\ref{fita})
is  substituted in Eq.~(\ref{fdglapmellin}),  the singular factor
$x^{-\alpha}$ affects the small -$x$ behavior of $g_1$ and changes
its asymptotics Eq.~(\ref{dglapas}) for $g_1$ for the Regge
asymptotics. Indeed, the small- $x$ asymptotics is governed by the
leading singularity $\omega = \alpha$, so
\begin{equation}\label{asdglap}
g_{1~DGLAP} \sim C(\alpha)(1/x)^{\alpha}\Big((\ln(Q^2/\Lambda^2))/
(\ln(\mu^2/\Lambda^2))\Big)^{\gamma(\alpha)}.
\end{equation}
 Obviously,  the actual DGLAP asympotics of Eq.~(\ref{asdglap}) is of the Regge type,
 and differs a lot from the conventional DGLAP asympotics of
 Eq.~(\ref{dglapas}). Indeed it 
 looks similar to
 our asymptotics given by Eq.~(\ref{gnsas}), namely by incorporating the singular
factors into DGLAP initial parton densities it leads to the steep rise of $g_1^{DGLAP}$ 
and therefore to a successful description of DGLAP at small $x$.
In Ref.~\cite{egt3} it is shown that without the singular factor
$x^{-\alpha}$ in the fit of Eq.~(\ref{fita}), DGLAP would not be
able to work successfully at $x \leq 0.05$. In other words, the
singular factors in DGLAP fits
 mimic the total resummation of logarithms of $x$ of
Eqs.~(\ref{gnsint},\ref{gnsas}).   To be more specific, although both (\ref{asdglap})
and (\ref{gnsas}) predict the Regge asymptotics for $g_1$, there
is a numerical difference in the intercepts: Eq.~(\ref{asdglap}) predicts
that the intercept of $g_1^{NS}$ should be $\alpha = 0.57$, a value 
which is greater than our predicted non-singlet intercept $ \Delta_{NS} =
0.42$. Therefore the non-singlet $g_1^{DGLAP}$ grows, when $x \to 0$, faster
than our predictions.   Such a rise however is too steep and 
contradicts the results obtained in Refs.~\cite{egt2} and
confirmed in Refs.~\cite{kat}. 

\section{Conclusions}

We have explicitly shown, by direct comparison of 
Eqs.~(\ref{dglapas}) and (\ref{asdglap}), 
 that the singular factor $x^{-\alpha}$ in the
Eq.~(\ref{fita}) for the initial quark density converts the
exponential DGLAP-asympotics into the Regge one. On the other
hand, comparison of Eqs.~(\ref{gnsas}) and (\ref{asdglap})
also shows that this singular factor in the DGLAP fits mimics the
total resummation of logarithms of $x$. This type of factors can be
dropped when the total resummation of logarithms of $x$ performed
in Ref.~\cite{egt2} is taken into account. The remaining terms, 
which are regular in 
$x$ in the DGLAP fits (the terms in squared brackets in
Eq.~(\ref{fita})), can obviously be simplified or even dropped at
small $x$ and replaced by constants. A more  detailed analysis as well 
as a suggestion to combine the leading logarithms resummation at 
small $x$ with DGLAP can be found in Ref.~\cite{egt3}.
The above results lead to an interesting
conclusion: the expressions for the initial parton densities  $\delta q$ 
used in DGLAP analyses have been commonly
believed to be related to non-perturbative QCD effects; indeed they
actually mimic the contributions of the perturbative QCD, so the
whole impact of the non-perturbative QCD effects on $g_1$ at small $x$ is
not large and can be approximated by a normalization constant.

\end{document}